\documentclass[%
reprint,
superscriptaddress,
 amsmath,amssymb,
pre,
]{revtex4-1}

\usepackage[english]{babel}
\usepackage[utf8x]{inputenc}
\usepackage[colorinlistoftodos]{todonotes}
\usepackage[T1]{fontenc}
\usepackage{mathtools}
\usepackage[per-mode=symbol]{siunitx}
\usepackage{soul}
\usepackage{multirow}
\usepackage{hyperref}

\usepackage{graphicx}
\usepackage{dcolumn}
\usepackage{bm}
\usepackage{hyperref}

\newcolumntype{P}[1]{>{\centering\arraybackslash}p{#1}}

\begin{document}

\preprint{APS/123-QED}


\title{Adiabatic lapse rate of non-ideal gases: The role of molecular interactions and vibrations}


\author{Bogar Díaz}
\email{bdiaz@iem.cfmac.csic.es}

\affiliation{Instituto de Estructura de la Materia, CSIC, Serrano 123, 28006 Madrid, Spain}
\affiliation{Departamento de Física de Altas Energías, Instituto de Ciencias Nucleares,
Universidad Nacional Autónoma de México, Apartado Postal 70-543, Ciudad de México, 04510, México} 

\author{J. E. Ram\'irez}
\email{jerc.fis@gmail.com}
\affiliation{Departamento de F\'isica de Part\'iculas and Instituto Galego de Física de Altas Enerxías, Universidad de Santiago de Compostela, E-15782 Santiago de Compostela, Spain}

\begin{abstract}

We report a formula for the dry adiabatic lapse rate that depends on the compressibility factor and the adiabatic curves. Then, to take into account the non-ideal behavior of the gases,  we consider molecules that can move, rotate, and vibrate and the information of molecular interactions through the virial coefficients. We deduce the compressibility factor in its virial expansion form and the adiabatic curves within the virial expansion up to any order. With this information and to illustrate the mentioned formula, we write the lapse rate for the ideal gas, and the virial expansion up to the second and third coefficient cases. To figure out the role of the virial coefficients and vibrations, under different atmospheric conditions, we calculate the lapse rate for Earth, Mars, Venus, Titan, and the exoplanet Gl 581d.  Furthermore, for each one we consider three models in the virial expansion: van der Waals, square-well, and hard-sphere.  Also, when possible, we compare our results to the experimental data. Finally, we remark that for Venus and Titan, which are under extreme conditions of pressure or temperature, our calculations are in good agreement with the observed values, in some instances.

\end{abstract}

\maketitle
\section{Introduction}

The \textit{lapse rate}, $\Gamma$, of an astronomic object's atmosphere is by definition the rate of change of its temperature with respect to height. It has been observed for many objects within the solar system with the help of space missions, for instance, cassini, Venera, Mariner, MESSENGER, rovers, Pioneer, Voyager, and others \cite{Kasprzak1990, Lindal1983, Mokhov2006, zurek_2017}.
One of the hot topics of astronomy is the search for habitable planets beyond Earth. A critical and necessary condition for a planet to be habitable is that the atmospheric conditions allow the existence of permanent liquid water on its surface \cite{astrobio,HuDi}. This is one of the reasons that render the study and the determination of the lapse rate  for other astronomical objects necessary.

The usual theoretical approach to study the lapse rate is the so-called dry adiabatic lapse rate (DALR). In this model, each parcel of the atmosphere is considered to be in thermal equilibrium and exchange no heat with its surroundings. Also, it makes use of the hydrostatic equation and the equation of state of the gas. The most elemental estimation of the DALR is obtained by considering the ideal gas model (denoted henceforth as $\Gamma^\text{IG}$), but this calculation yields a value that is far from experimental data \cite{Catling2015}.  This is expected, on one hand, since the atmospheres are composed of many other elements (Earth's atmosphere contains traces of vapor, for example). On the other hand, the molecules of a \textit{real} gas can vibrate and interact between them (non-ideal gas behavior). 

Some authors try to incorporate these non-ideal gas effects of the molecules using experimental information about the gas in the formula of $\Gamma^\text{IG}$ \cite{Catling2015, Vallero2014}. This is an incorrect procedure because the $\Gamma^\text{IG}$ strongly depends on the initial assumptions. A derivation of a DALR for real gases is found in Ref.~\cite{Staley1970}, where the author applied his formula for Venus, obtaining a good approximation to the experimental value. 
A limitation of this approach is the availability of experimental data in the range of atmospheric conditions for other astronomic objects. From the theoretical point of view, a shortcoming of this method is the impossibility of quantifying the origin of the correction, namely, the molecular vibrations, or the molecular interactions, or both. 

An attempt to quantify the contribution of molecular interactions can be found in Ref.~\cite{JHON}. There, the authors analyze the effect of the second virial coefficient on the DALR. This means that they include the possibility of two-particle interactions only (by $n$-particle interactions, we mean that the interaction potential takes place between $n$ particles). This is quite restrictive; in fact, in some instances, the $n$-particle interactions or molecular vibrations could be relevant. Nevertheless, they find that in Titan, given its atmospheric conditions, these interactions have a strong effect on the DALR towards the observed value. The purpose of this paper is to incorporate the information of $n$-particle interactions and the possibility of  molecular vibrations to the DALR model.

The plan of the paper is the following. In Sec. \ref{sect: 2} we deduce a formula for the DALR that depends on the compressibility factor (for fluids) and the adiabatic curves. Next, from statistical mechanics, in Sec. \ref{section:ES} we deduce the compressibility factor and the adiabatic curves within the virial expansion up to any order and allowing molecular vibrations. In Sec. \ref{sec:pcdalr}, we combine the results of Secs. \ref{sect: 2} and \ref{section:ES} to compute the DALR for some particular cases of the virial expansion for vibrating and non-vibrating molecules. There, we also discuss three instances of virial coefficients related to simple fluid models, namely, van der Waals, square-well, and hard-sphere. As an application of the formulas obtained, we devote Sec. \ref{results} to calculate the DALR for Earth, Mars, Venus, Titan, and the exoplanet Gl 581d. Finally, in Sec. \ref{sect: Dandc} we compare our results to the observational data and in the case of Venus to those reported in Ref. \cite{Staley1970}, and we present our conclusions.

\section{Dry adiabatic lapse rate} \label{sect: 2}
In this section, we shall determine a general expression for the DALR for non-ideal gases.
The mathematical definition of the lapse rate is
\begin{equation}\label{eq:LRdefinition}
    \Gamma \coloneqq -\frac{\mathrm{d}T}{\mathrm{d} z}\,,
\end{equation}
where $T$ is the temperature and $z$ is the height.
We show below that it turns out to be proportional to the well-known DALR of monocomponent ideal gases given by
\begin{align}\label{eq:lrig}
\Gamma^\text{IG}= \frac{M_\text{mol}g}{C^{IG}_P} \,,
\end{align}
where $M_\text{mol}$ is the molar mass of the gas of the atmosphere, $g$ denotes the magnitude of the acceleration due to gravity close to the surface of the astronomical object, and $C^{IG}_P$ is the specific heat at constant pressure for ideal gases, the value of which is $C^{IG}_P=\frac{5+f_r}{2} R$, where $f_r=0$ , 2 and 3, for monoatomic, diatomic or linear, and polyatomic molecules, respectively \cite{schroeder2000}. $R$ is the gas constant ($R=8.31$ \si{\joule\per\mole\per\kelvin}). 
Notice that, as is usual in the literature, we take $g$ as a constant even when it depends on the height (this can be easily incorporated \footnote{Using Newton's law of universal gravitation the acceleration due to gravity as a function of the high $z$ can be written as $g(z)= g / \left(1+ z/r \right)^2$, where $r$ denotes the radio of the astronomic object.}). The justification is that, in the examples, the variation of the value of $g$ from the surface to the top of the troposphere is less than $1\%$. In the equation above and in what follows, we consider $1$ mole of gas.

We start our derivation from the definition of the compressibility factor, $Z$,
 \begin{equation}\label{eq:CF}
    Z \coloneqq \frac{PV}{RT} \,,
\end{equation}
where $P$ is the pressure, $V$ is the volume, and $T$ is the absolute temperature of the gas \cite{mcquarriethermo}. The value of $Z$ can be determined experimentally (see, for example, Ref.~\cite{dataZ}), obtaining then an equation of state that describes \textit{real} gases. As we mentioned, a shortcoming of this approach is the availability of data for the conditions of interest in the astronomical objects under study. As an alternative, there exist theoretical models that propose some specific functions for $Z$, that can be physically interpreted, such as ideal gas  ($Z=1$), van der Waals \cite{cengel}, and Redlich-Kwong \cite{mcquarriethermo} models, among others. Within the theoretical approaches, we are interested in the \textit{virial expansion} \cite{mcquarrie}, which accounts for interactions between successively larger groups of molecules. For completeness, we derive the corresponding $Z$ in Sec.~\ref{section:ES} from statistical mechanics. At this moment, it is sufficient to know that in this approach $Z$ can be expressed as a function that only depends on $V$ and $T$, i.e., $Z= Z(V, T)$.
  
Furthermore, we are interested in the analysis of the adiabatic lapse rate. The adiabatic process dictates the specific forms of the curves in the different diagrams, which are called adiabatic curves. We obtain these curves in Sec.~\ref{section:ES}. For the general formula that we are after, it is enough to use the fact that, in the region we are interested in, the volume can be written as a function of the temperature, $V=V(T)$, on each adiabatic curve. From now on, for functions that only depend on $T$ we denote with a prime its derivative with respect to the temperature, for example, $V'$.

According to the general considerations discussed above, from \eqref{eq:CF} we have
\begin{align}
    \mathrm{d}P=\frac{R}{V}\left(Z+T\frac{\partial Z}{\partial T}-ZT\frac{V'}{V}+T \frac{\partial Z}{\partial V} V'  \right) \mathrm{d}T\,.
\label{eq:dPZ}
\end{align}
On the other hand, using the fact that the density of the gas, $\rho$, in the atmosphere is given by $\rho=M_\text{mol}/V$ we can write the hydrostatic equation as
\begin{align}\label{eq:Hydr}
  \mathrm{d} P=- \frac{M_\text{mol}g}{V}   \mathrm{d} z  \,.
\end{align}
Substituting \eqref{eq:Hydr} in \eqref{eq:dPZ} and taking into account \eqref{eq:LRdefinition} and \eqref{eq:lrig}, we obtain
\begin{equation}
    \Gamma= \Gamma^\text{IG} \frac{C^{IG}_P}{R}\left(Z+T\frac{\partial Z}{\partial T}-ZT\frac{V'}{V}+T \frac{\partial Z}{\partial V} V'  \right)^{-1} \,.
\label{eq:LR}
\end{equation}

Let us make some remarks about Eq. \eqref{eq:LR}. 

(1) Regarding the compressibility factor, this equation only makes use of the fact that $Z$ can be expressed as a function of the volume and temperature. This happens in the virial expansion to any order and this is also true for the compressibility factor obtained from other equations of state, for example, the van der Waals equation.

(2) As we have mentioned before, in the virial expansion $Z$ accounts for interactions between successively larger groups of molecules. However, by itself, it is not sensitive to the vibrational state of the molecules. \label{novibra}

(3)  We must emphasize that this equation is evaluated on an adiabatic curve $V=V(T)$. We show in Sec.~\ref{section:ES} that the energy, which is used to derive the adiabatic curves, is not only modified by the virial coefficients, but it also takes into account the contribution of molecular vibrations. \label{sivibra}

(4) Of course, Eq. \eqref{eq:LR} reproduces \eqref{eq:lrig}, because for the ideal gas case $Z=1$, and the adiabatic curves of ideal gases satisfy $V'/V=-C^{IG}_V/\left(RT\right)$, where $C^{IG}_V$ is the specific heat at constant volume for ideal gases, which fulfills $C^{IG}_V =C^{IG}_P-R$ ($=\frac{3+f_r}{2} R$). Using this, we have $\Gamma= \Gamma^\text{IG}$. \label{ideal case}

(5) Notice that, so far, the consideration of a monocomponent gas in \eqref{eq:LR} is only explicit in $\Gamma^\text{IG}$ and $C^{IG}_P$ which can be generalized to the case of multicomponent gases. However, in that scenario the compressibility factor and the adiabatic curves are modified in a non-trivial way. In what follows, we restrict ourselves to the monocomponent case, which suffices to understand the atmospheric features of the astronomical objects that we are interested in.

\section{Equation of state and adiabatic curves} \label{section:ES}

We start from the partition function for the non-ideal gases in which we are interested. This allows us to compute the average pressure and the average energy. Then, using the ensemble postulate of Gibbs, we obtain the thermodynamic variables of interest, the pressure and the internal energy. From the pressure expression, we derive the equation of state for this system. Then, we identify the compressibility factor, which corresponds to the well-known functional expression of $Z$ in the virial expansion. On the other hand, we need the internal energy to describe the adiabatic processes, i.e., the adiabatic curves, which we calculate at the end of this section.

The partition function, in Mayer's representation, of a gas constituted by $N$ indistinguishable particles is given by \cite{Mayer1958}
\begin{align}
    Q= Q_\text{MI} \frac{\left( q_\text{trans} q_\text{rot} q_\text{vib} \right)^N}{N!}\,, \label{eq:pfO}
\end{align}
where $Q_\text{MI}$ is the partition function that codifies the molecular interactions, while the other factors are the partition functions corresponding to the fact that the particles move in a three-dimensional space ($q_\text{trans}$),  rotate ($q_\text{rot}$), and vibrate ($q_\text{vib}$). These partition functions are given by \cite{mcquarrie,Mayer1958,mayerVirial}
\begin{subequations}
\begin{align}
    Q_\text{MI}&= \exp\left( -N \sum_{k=1}\frac{B_{k+1}}{kV^k}  \right) \,, \label{eq:pf1}\\
    q_\text{trans}&= \left( \frac{2\pi M k_B T}{h^2} \right)^{3/2}V  \,, \label{eq:pf2}\\
    q_\text{rot}&=  \frac{T^{f_{r}/2}}{\theta_\text{rot}} \,,  \label{eq:pf3}\\
    q_\text{vib}&=  \prod_{j=1}^m \frac{\exp(-\theta_j/2T)}{1-\exp(-\theta_j/T)} \,,  \label{eq:pf4}
\end{align}
\end{subequations}
where \eqref{eq:pf1} is valid only at gaseous regimes up to the saturation point \cite{ushcats1, ushcats2, ushcats3} and the $B_{k+1}$ are the so-called virial coefficients, which generically are sums of the cluster integrals involving the $(k+1)$-particle interactions \cite{mayerbook}. 
They are, by construction, functions that can \textit{only} depend on the temperature $T$ \cite{mcquarriethermo} 
(this statement becomes invalid at the vicinity of the boiling point \cite{ushcats4}). 
In \eqref{eq:pf2}, $M$, $k_B$, and $h$ are the mass of the molecule, the Boltzmann constant, and the Planck constant, respectively. In \eqref{eq:pf3}, $\theta_\text{rot}$ is a constant related to the characteristic rotational temperatures. Finally, in \eqref{eq:pf4}, $m$ is the number of natural vibrational frequencies, $\nu_j$, and $\theta_j \coloneqq h\nu_j/k_B$ are known as vibrational temperatures \cite{mcquarrie}. Concerning $q_\text{vib}$ notice that (a) its value is gas-dependent and (b) for models where there is no need to incorporate the molecular vibration information it is enough to set $q_\text{vib}=1$. This allows us to turn off the vibrational modes, and (c) it is a function that only depends on $T$.

Now, the average pressure and average energy, in the canonical ensemble, are calculated as \cite{mcquarrie}
\begin{subequations} \label{defPE}
\begin{align}
    \langle P \rangle=k_B T\left(\frac{\partial }{\partial V}\ln Q \right)_{N,T}\,, \\
    \langle E \rangle=k_B T^2 \left(\frac{\partial }{\partial T}\ln Q \right)_{N,V}\,.
\end{align}
\end{subequations}
Therefore, using \eqref{defPE} and the partition function \eqref{eq:pfO}, for $1$ mole, we obtain 
\begin{subequations}
\begin{align}
\frac{\langle P \rangle V}{RT}&=1+\sum_{k=1}\frac{B_{k+1}}{V^k}\,, \label{EoE}\\
\langle E \rangle & =C^{IG}_VT+RT^2\Phi-RT^2 \sum_{k=1}\frac{B'_{k+1}}{kV^k}\,,\label{EI}
\end{align}
\end{subequations}
where $\Phi \coloneqq q'_\text{vib}/q_\text{vib}$. Furthermore, through the ensemble postulate of Gibbs, we have that the average pressure and average energy coincide with the pressure and internal energy of the system in the thermodynamic context, i.e.,  $\langle P \rangle=P$ and $\langle  E \rangle =U$. 
Using the latter result, \eqref{EoE}, and comparing with \eqref{eq:CF}, we have that
\begin{align}
Z= 1+\sum_{k=1}\frac{B_{k+1}}{V^k}  \,. \label{eq:VE}
\end{align} 
In this way, we obtain the well-known virial expression for the compressibility factor $Z$.
This is the function needed in \eqref{eq:LR} to compute the lapse rate. 

Notice that the expression of $Z$ confirms remark 2, 
which says that the vibrations do not modify the equation of state and that the energy \eqref{EI} takes into account not only the molecular interactions but also the vibrational state of the molecules (through  $\Phi$), which confirms the claim we made in remark 3. 

Additionally, for \eqref{eq:LR} we need the adiabatic curves. From the first law of thermodynamics, if we consider an adiabatic process we have
\begin{align}
    dU+PdV=0\,. \label{eq:ap}
\end{align}
Plugging the equation of state \eqref{EoE} and the energy \eqref{EI} into \eqref{eq:ap}, we obtain that the adiabatic curves in the $V$-$T$ diagram are given by
\begin{align}
    T\left[Vq_\text{vib}
    \exp\left(T\Phi- \sum_{k=1} \frac{\mathcal{B}_{k+1}}{kV^k} \right) \right]^{R/C_V^{IG}}=\varepsilon_0 \,,
    \label{eq:e0}
\end{align}
where $\mathcal{B}_{k+1} \coloneqq TB'_{k+1}+B_{k+1}$ and $\varepsilon_0$ is a constant, that can be determined by the atmospheric conditions on the surface of the astronomical object. However, to be able to use \eqref{eq:e0} in \eqref{eq:LR}, we need to solve for the volume in terms of the temperature. This last step can not be analytically done for an arbitrary order of the virial expansion. We remark that it strongly depends on the order considered of the virial expansion and that, on the other hand, the vibrations pose no difficulty. 

\section{Physical models and their DALR} \label{sec:pcdalr}

Here, we discuss the following models: ideal gas, the virial expansion up to the second coefficient (both with and without vibrations), and the virial expansion up to the third coefficient with vibrations. We use these models in Sec. \ref{results} to calculate the DALR for some astronomical objects and compare the results with the available observations.

\subsection{Ideal gas with vibrational modes}\label{IG+v}

Let us consider the case of an ideal gas and incorporate the effect of vibrational modes. Physically, this case represents a gas composed of non-interacting molecules that can move, rotate, and vibrate. Under these circumstances $B_k=0$ for all $k$. Then \eqref{eq:e0} reduces to
\begin{align}
TV^{{R/C^{IG}_V}}[q_\text{vib} \exp(T\Phi )]^{{R/C^{IG}_V}}=\varepsilon_0\,.
\end{align}
The volume as a function of the temperature is 
\begin{align}
V=\left(\frac{\varepsilon_0}{T} \right)^{ C^{IG}_V/R}\frac{\exp(-T\Phi)}{q_\text{vib}}\,. \label{volt}
\end{align}
Then,
\begin{align}
\frac{V'}{V}=-\frac{ C^{IG}_V}{RT}-T\Phi'-2\Phi \, . \label{eq: curigv}
\end{align}
Using \eqref{volt}, \eqref{eq: curigv}, and $Z=1$ in \eqref{eq:LR}, we obtain that the DALR is given by
\begin{align}
   \Gamma=\Gamma^\text{IG} \frac{C^{IG}_P}{C^{IG}_P+ RT^2\Phi'+2RT\Phi}\,. \label{dalricv}
\end{align}
Notice that the corresponding value of $\varepsilon_0$ is not needed in \eqref{dalricv}. 

\subsubsection*{Ideal gas case without vibrations}

As we have mentioned, if we do not want to consider the contributions of molecular vibrations it suffices to set $q_\text{vib}=1$, then $\Phi=0$. Using this in \eqref{eq: curigv} and \eqref{dalricv}, we find
\begin{subequations}
\begin{align}
\frac{V'}{V}&=  - \frac{ C^{IG}_V}{RT} \,, \label{cic} \\
\Gamma&=\Gamma^\text{IG}\,.
\end{align}
\end{subequations}
 This proves the claim we made in remark 4.

\subsection{Virial expansion up to second order including vibrations} \label{IG+B+V}

Now, we add to the case in Sec. \ref{IG+v} the possibility of two-particle interactions. Mathematically, this means we consider the virial expansion up to second order, i.e., $B_k=0$ for $k>2$, therefore \eqref{eq:e0} reduces to
\begin{align}
    T\left[V \exp\left(- \frac{\mathcal{B}_2}{V} \right)q_\text{vib} \exp\left(T\Phi \right)\right]^{R/C^{IG}_V}=\varepsilon_0\, ,
\end{align}
from which
\begin{align}
V(T)=\frac{\mathcal{B}_2}{W(x)}\,, \label{eq:ad-Bvib}
\end{align}
where $x\coloneqq\mathcal{B}_2 q_\text{vib} \exp(T\Phi)\left(\frac{T}{\varepsilon_0}\right)^{C^{IG}_V/R}$ and $W(x)$ is the Lambert function \footnote{The Lambert function $W$ is defined by $W(z) \exp\left(  W(z) \right) = z$ and it cannot be expressed in terms of elementary functions.}.

Now, using $Z= 1+B_2/V$ and \eqref{eq:ad-Bvib} in \eqref{eq:LR}, we obtain the following DALR: 
\begin{widetext}
\begin{align}
\Gamma =  \Gamma^\text{IG}  \frac{C^{IG}_P}{R} \left\{1+W(x)-T\left[1+\frac{2B_2 W(x)}{\mathcal{B}_2}  \right] \left[\frac{\mathcal{B}_2'}{\mathcal{B}_2}-\frac{1}{1+W(x)}\frac{x'}{x}  \right] \right\}^{-1}\,, \label{dalr b vib}
\end{align}
\end{widetext}
where the quotient $x'/x$ is explicitly given by
\begin{align}
\frac{x'}{x}=\frac{\mathcal{B}'_2}{\mathcal{B}_2}+\frac{ C^{IG}_{V}}{RT}+2\Phi+T\Phi'\,. \label{eq: x prima entre x}
\end{align}
Notice that the corresponding value of $\varepsilon_0$ \textit{is} needed in \eqref{dalr b vib} because $x$ depends on it. 
\subsubsection*{Virial expansion up to second order without vibrations}\label{IG+B}
We can directly obtain the case without vibrations from the analysis in Sec. \ref{IG+B+V}. The formula for the DALR looks like \eqref{dalr b vib}. By setting $q_\text{vib}=1$, now we have
\begin{subequations}
\begin{align}
x &=\mathcal{B}_2 \left(\frac{T}{\varepsilon_0}\right)^{C^{IG}_V/R}\,, \label{sx} \\
\frac{x'}{x}&=\frac{\mathcal{B}_2'}{\mathcal{B}_2}+\frac{ C^{IG}_{V}}{RT}\,.
\end{align}
\end{subequations}

It is worth mentioning that this case was studied in Ref. \cite{JHON} by the authors and collaborators. There, the following formula for the DALR was obtained
\begin{align}
     \Gamma= \frac{M_\text{mol}g}{2B_2  P' } \left( \sqrt{1+\frac{4P }{RT}B_2} -1 \right) \,, \label{eq: previous dalr}
\end{align}
where $P$ is the pressure on the adiabatic curve. Equation \eqref{eq: previous dalr} looks different from \eqref{dalr b vib}; the reason is that in Ref. \cite{JHON} the adiabatic curves were used in the $P$-$T$ diagram. The equivalence between \eqref{dalr b vib} and \eqref{eq: previous dalr} can be proved in the following way. We need the equation of state over the adiabatic curve, this is,
\begin{align}
   \left( \frac{P}{RT} \right) \left(  \frac{\mathcal{B}_2}{W(x)} \right) &=1+B_{2} \left(  \frac{W(x)}{\mathcal{B}_2}  \right)\,, \label{esb}
\end{align}
We have used \eqref{eq:ad-Bvib}, with $x$ given by \eqref{sx}, to substitute the volume. From \eqref{esb}, we obtain $P$ and $P'$, and plugging them into \eqref{eq: previous dalr} we obtain \eqref{dalr b vib}.

We believe that the derivation presented here is conceptually clearer and shorter than the one appearing in Ref. \cite{JHON}. Moreover, here we have obtained a formula for the DALR to any order in the virial expansion, not only up to the second one as in Ref. \cite{JHON}.

\subsection{Third order virial expansion including vibrations}\label{subnume}

The last model that we want to analyze includes also the possibility of three-particle interactions.
This case corresponds to the third order virial expansion, $B_k=0$ for $k>3$. For the adiabatic curves, \eqref{eq:e0} reduces to
\begin{align}
    T\left[V \exp\left(- \frac{\mathcal{B}_2}{V} - \frac{\mathcal{B}_3}{2V^2} \right)q_\text{vib} \exp\left(T\Phi \right)\right]^{ \frac{R}{C^{IG}_V}}=\varepsilon_0\,. \label{adt}
\end{align}
Unfortunately, from \eqref{adt} we cannot write the volume as a function of the temperature in closed form. However, we can use numerical methods to obtain the volume in the region of interest. To be precise, we search for a solution of \eqref{adt} using Newton methods (starting in the ideal gas volume with or without vibration, accordingly), starting with the temperature at the surface of the astronomical object and decreasing it in steps of $\Delta T=\SI{0.01}{\kelvin}$ until reaching the value of the temperature that corresponds to the highest part of the troposphere. 
Moreover, the derivative is computed by using the five-point stencil method (with a spacing between points of $0.001$). Both the volume and its derivative are interpolated with cubic splines. 

Finally, under these circumstances and using $Z=1+B_2/V+B_3/V^2$, the DALR is given by
\begin{align}
    \Gamma=&\Gamma^\text{IG} \frac{C^{IG}_P}{R}\left[1+\frac{\mathcal{B}_2}{V}+\frac{\mathcal{B}_3}{V^2}\right. \nonumber\\
      & \left. -\frac{T V'}{V}\left(1+\frac{2B_2}{V}+\frac{3B_3}{V^2}  \right) \right]^{-1}\,. \label{dalr b_3 +vib}
\end{align}

\subsection{Virial coefficients}

As we have mentioned $B_2$ and $B_3$ take into account two-particle and three-particle interactions, respectively. Their explicit functional form is dictated by the molecular interaction model. For the analysis of the lapse rate in the astronomical objects of interest, we have chosen three models: van der Waals, square-well, and hard-sphere. Their second virial coefficients are given by
\begin{subequations} \label{B_2}
\begin{align}
B_2^\text{vdW}&= a+\frac{b}{T}\,, \label{vdw}\\
B_2^\text{sw}&= \tilde{a}[1-(d^3-1)f]\,,\label{eq:B2SW}\\
B_2^\text{hs}&= \tilde{a}\,, \label{hseq}
\end{align}
\end{subequations}
respectively \cite{vcHS,vcHS2,vcSW,vcSW2}. The function $f$ in \eqref{eq:B2SW} is $f=\exp(c/T)-1$. The parameters $a$, $\tilde{a}$, $b$, $c$, and $d$ have a physical interpretation and their values are gas dependent. As it is well known, the van der Waals model takes into account the volume of the molecules and the molecular interactions. In \eqref{vdw}, $a$ represents the average excluded volume and $b$ is associated with the attractive interaction. The square-well model also considers the volume of the molecules and an attractive interaction. 
In \eqref{eq:B2SW}, $\tilde{a}$ is the volume of the molecules, considered as hard spheres, while $d$ and $c$ are related to the range and amplitude of the attractive interaction potential. Therefore, the square well is a generalization of the hard-sphere model, which only considers the volume $\tilde{a}$ in \eqref{hseq}.

Using the parameters introduced in \eqref{B_2}, we can write the third virial coefficient \cite{vcHS,vcHS2,vcSW,vcSW2} as 
\begin{subequations} \label{B_3}
\begin{align}
B_3^\text{vdW}&=    a^2 \,, \label{vdwthird}\\
B_3^\text{sw}&=     \frac{\tilde{a}^2}{8}[5-(d^6-18d^4+32d^3-15)f   \nonumber \\
  &+(-2d^6+36d^4-32d^3-18d^2+16)f^2   \label{eq:B3SW} \\
  &-(6d^6-18d^4+18d^2-6)f^3]   \,\,  \, \textrm{if} \, d<2 \,, \nonumber \\  
B_3^\text{hs}&=     \frac{5\tilde{a}^2}{8} \,. \label{hsthird}
\end{align}
\end{subequations}
Notice that \eqref{eq:B3SW} is valid only if $d<2$; for the case $d \geq 2$ the corresponding formula is reported in Ref. \cite{vcSW}. For the gases that we consider below \eqref{eq:B3SW} suffices.

\begin{figure}[ht!]
\centering
\includegraphics[scale=1]{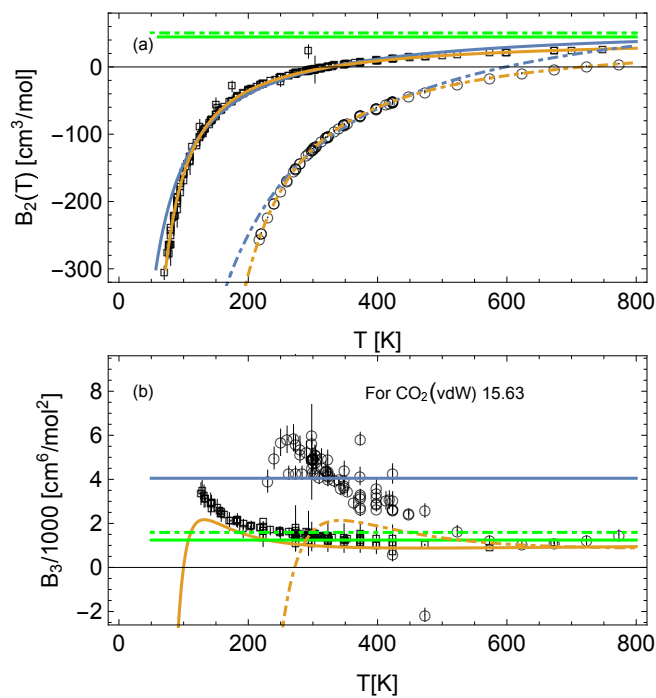}
\caption{Experimental data (points) of the virial coefficients $B_2$ (a) and $B_3$ (b) for N$_2$ (squares) and CO$_2$ (circles). In (a), blue (gray) and yellow (medium gray) lines correspond to the curve fits to the $B_2$ data for the van der Waals and square-well models. Green (light gray) lines are the $B_2$ value for the hard-sphere interaction. Solid and dot-dashed lines denote the information about the gases N$_2$ and CO$_2$, correspondingly.
In (b), color lines are the closed-form expressions of $B_3$ in \eqref{B_3} using the corresponding fit parameters of $B_2$.}
\label{fig:virial}
\end{figure}

The parameters that appear in \eqref{B_2} can be obtained by fitting these equations to the experimental data for the second virial coefficient of the gas of interest. In our case, we are interested in N$_2$ and CO$_2$ (see Sec. \ref{results}). We show in Table~\ref{tab:virial} the values obtained for the parameters and in Fig.~\ref{fig:virial} the curves fitting the experimental data (see Ref.~\cite{dymond2002virial}) for the three models considered.

\begin{table}[ht!] \centering
    \caption{Values of the parameters in \eqref{B_2} obtained by fitting these equations to the experimental data for the second virial coefficient.}
    \label{tab:virial}
    
    \begin{tabular}{| c| c| c| c|}
 \hline
 Model & Parameters & N$_2$ & CO$_2$\\
 \hline
  \multirow{2}{*}{vdW} & $a$ [$\SI{}{\cubic\centi\meter\per\mole}$] & $63.6\pm0.9 $&$125\pm3 $ \\
   & $b$ [$\SI{}{\cubic\centi\meter\kelvin\per\mole}$] & $-20786\pm266 $ & $-74780\pm1281 $ \\
 \hline
\multirow{3}{*}{sw} & $\tilde{a}$ [$\SI{}{\cubic\centi\meter\per\mole}$] & $44.5\pm0.2$ &$50.4\pm0.3$ \\
  & $d$ & $1.619\pm0.003$ & $1.400\pm0.002$ \\
  & $c$ [$\SI{}{\kelvin}$] & $87.8\pm0.7$& $324\pm2$\\
 \hline
 hs & $\tilde{a}$ [$\SI{}{\cubic\centi\meter\per\mole}$] & $44.5 \pm 0.2$& $50.4 \pm 0.3 $\\
 \hline
    \end{tabular}
\end{table}

\section{Results} \label{results}

Here, we present an application of the formulas developed in Sec. \ref{sec:pcdalr}. As we want to illustrate the role of the virial coefficients and vibrations, we pick out astronomical objects having diverse atmospheric conditions. Our selection is Earth, Mars, Venus, Titan, and the exoplanet Gl 581d. In Table~\ref{tab:table1} we present the relevant information for each astronomical object, namely, the most abundant gas in its atmosphere (major constituent), $\Gamma^\text{IG}$; the experimental lapse rate (denoted by $\Gamma^\text{Obs}$), the atmospheric conditions on the surface (for the exoplanet Gl 581d there is a wide range of pressures and temperatures allowing the presence of liquid water \cite{HuDi, Words, von}); and their corresponding $g$ value. In the computation of the DALR, the atmosphere constitution is considered as monocomponent, composed of the most abundant gas. Remember, for the astronomical objects under discussion, theses gases are N$_2$ and CO$_2$. The microscopic information required for the calculations is the following: the rotational degrees of freedom are $f_r=2$ due to the linearity of these molecules. For the vibrational temperatures, $\theta_j$, we use the wave numbers: (i) for CO$_2$, $\SI{667.3}{\cm}^{-1}$ (with degeneracy 2), $\SI{1341.5}{\cm}^{-1}$ and $\SI{2349.3}{\cm}^{-1}$ \cite{dataCO2}, and (ii) for N$_2$, $\SI{2328.72}{\cm}^{-1}$ \cite{dataN2}.

\begin{table*}[ht!] 
    \caption{Information about the astronomical objects under consideration.}
    \label{tab:table1}
    
    \begin{tabular}{|P{1.9cm}|P{1.6cm}|c|P{1.5cm}|P{1.5cm}|P{1.5cm}|P{1.5cm}| P{1.5cm}|}
 \hline 
      Astronomical object &	Major constituent &	Composition	[\%]& $\Gamma^{\text{IG}}$ [$\SI{}{\kelvin\per\kilo\meter}$]&	$\Gamma^{\text{Obs}}$ [$\SI{}{\kelvin\per\kilo\meter}$] & $P_0$ [$\SI{}{\kilo\pascal}$] &$T_0$ [$\SI{}{\kelvin}$]& $g$ [$\SI{}{\meter\per\second^2}$] \\
\hline
Earth &	N$_2$	& 78    	& 9.44	& 6.5  &	101&	288& 9.80 \\
Titan &	N$_2$	& 94.2  	& 1.30 &	1.38 &	150&	94& 1.35\\
Mars & CO$_2$ & 96   & 5.61 & 2.5 &0.6	&215& 3.71\\
Venus &	CO$_2$	& 96.5 	& 13.42	& 8.4 &	9200&	737& 8.87\\
Gl 581d (A) & CO$_2$ & 96    & 30.70 & ---& 100 & 217 &20.30\\
Gl 581d (B) & CO$_2$ & 96    & 30.70 & ---& 2000 & 343 &20.30\\
Gl 581d (C) & CO$_2$ & 96    & 30.70 & ---& 5000 & 375 &20.30\\
\hline
    \end{tabular}
\end{table*}

In order to improve notation, we will denote the different models in Sec. \ref{sec:pcdalr} as follows: ideal gases with (without) vibrations by IG$+$vib (IG), virial expansion up to the second order with (without) vibrations by $B_2+\textrm{vib}$ ($B_2$), and virial expansion up to the third order with vibrations by $B_3+\textrm{vib}$.

\subsection{Adiabatic curves} \label{adiabacurvas}

The adiabatic curves, given in the general case by \eqref{eq:e0}, require the information of at least one point lying on the curve to calculate the constant $\varepsilon_0$. We choose to take the atmospheric conditions at the surface of the astronomical object in question. This information is provided in Table \ref{tab:table1} in terms of the temperature $T_0$ and the pressure $P_0$, but \eqref{eq:e0} is defined in the $V-T$ diagram. For our calculations, it is necessary to compute the corresponding volume of 1 mole of the gas under these pressure and temperature conditions. To do this, we substitute $T_0$ and $P_0$ in the equation of state and then solve for $V$. In particular, for the virial expansion up to $B_2$ and $B_3$, we obtain two and three $V_0$ values, respectively. Then, we discard complex or negative solutions, and also those that do not reduce to the ideal gas case when $B_2$ and $B_3$ are equal to zero and there are no vibrations. In Table \ref{tab:e0}, we summarize the values of $\varepsilon_0$ for all the models discussed in Sec. \ref{sec:pcdalr}.

\begin{table*}[ht!] \centering
    \caption{Values of $\varepsilon_0$.}
    \label{tab:e0}
    
    \begin{tabular}{|c| c| c|c| c| c|c| c| c|c| c| c|}
\hline
 & \multirow{2}{*}{IG} & \multirow{2}{*}{IG$+$vib} & \multicolumn{3}{c|}{van der Waals} & \multicolumn{3}{c|}{Square well} & \multicolumn{3}{c|}{Hard sphere}\\ \cline{4-12}
    &	&	&	$B_2$&	$B_2$+\textrm{vib}&	$B_3$+\textrm{vib}&	$B_2$&	$B_2$+\textrm{vib}&	$B_3$+\textrm{vib}&	$B_2$&	$B_2$+\textrm{vib}&	$B_3$+\textrm{vib}\\
\hline
Earth&	64.3729&	64.4586&	64.2943&	64.2971&	64.2973&	64.3076&	64.3104&	65.2783&	64.3729&	64.3758&	65.5306\\
Titan&	11.5267&	11.5267&	11.3274&	11.3274&	11.3799&	11.2154&	11.2154&	11.2654&	11.5265&	11.5265&	11.5783\\
Mars&	332.655&	350.177&	332.639&	350.16&	350.16&	332.628&	350.148&	350.148&	332.655&	350.177&	350.177\\
Venus&	39.742&	86.9272&	37.4769&	81.9728&	82.5694&	38.3444&	83.8704&	83.9243&	39.7043&	86.8447&	86.8916\\
Gl 581d (A)&	43.5396&	45.9154&	43.2026&	45.56&	45.5605&	42.9642&	45.3086&	45.3025&	43.5395&	45.9154&	45.9154\\
Gl 581d (B)&	24.9368&	30.6176&	23.3244&	28.6379&	28.7145&	23.3532&	28.6732&	28.684&	24.9311&	30.6106&	30.6147\\
Gl 581d (C)&	19.5838&	25.1675&	16.8087&	21.6012&	22.07&	17.0629&	21.928&	22.0244&	19.5626&	25.1403&	25.1557\\
\hline
    \end{tabular}
\end{table*}

\begin{figure*}[ht!]
\centering
\includegraphics[scale=1]{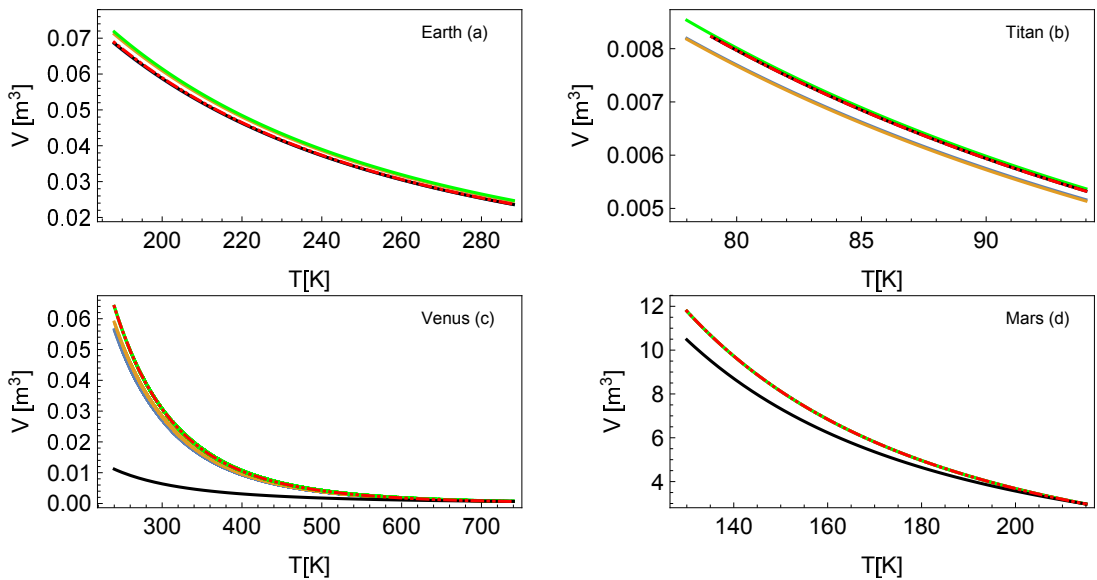}
\caption{Adiabatic curves obtained by considering the third order virial expansion and molecular vibrations for Earth (a), Titan (b), Venus (c), and Mars (d). The colors of the lines are the same as in Fig.~\ref{fig:virial}. The ideal gas prediction with vibrations and without vibrations are represented by a black solid and a red (gray) dot-dot-dashed lines, respectively.}
\label{fig:adcurves}
\end{figure*}

\begin{figure}[ht!]
\centering
\includegraphics[scale=1]{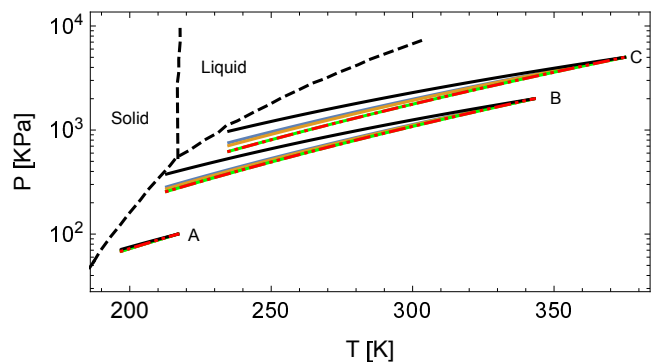}
\caption{Adiabatic curves considering the virial expansion up to third order with molecular vibrations and ideal gas prediction with and without vibrations for the exoplanet Gl 581d, for the three atmospheric conditions appearing in Table~\ref{tab:table1}. Lines and colors are the same as Fig. \ref{fig:adcurves}. The dashed black lines correspond to the phase diagram of CO$_2$ \cite{Span}}.
\label{fig:adcurves-exo}
\end{figure}

As an example, and to illustrate the effect of the interactions and vibrations, we show in Fig. \ref{fig:adcurves} the adiabatic curves for the third order virial expansion with vibrations together with the IG and IG$+$vib cases for Earth, Titan, Mars, and Venus. Those of the exoplanet Gl 581d, under the three atmospheric conditions shown in Table.~\ref{tab:table1}, are depicted in Fig.~\ref{fig:adcurves-exo}. Let us make some comments.

(a) The major deviation due to the molecular vibrations occurs in Venus.  This is to be expected as a consequence of the high atmospheric temperatures.

(b) On the other hand,  the deviations on Titan are primarily due to the molecular interactions, since its atmospheric conditions are close to the boiling point of N$_2$ (ca. $\SI{77}{\kelvin}$ at $\SI{101}{\kilo\pascal}$).

(c) The different atmospheric conditions in the exoplanet are such to avoid phase transitions, as we illustrate in Fig.~\ref{fig:adcurves-exo}. 



\subsection{Lapse rate}\label{sublr}

Here, we show the DALR under the conditions described in Sec. \ref{sec:pcdalr}, together the values obtained in Sec. \ref{adiabacurvas}, for the astronomical objects in Table \ref{tab:table1}.

Notice that the DALR is a function of the temperature $T$ only. It is possible to express $\Gamma$ in terms of the height $z$. To do this, we solve the differential equation for the corresponding DALR by the fourth order Runge-Kutta  method. In this way, we write $T=T(z)$, and thus $\Gamma=\Gamma(T(z))$.

In order to enhance the exposition, we divide the presentation into three groups: astronomical objects with atmospheres close to ideal gases (Earth and Mars), atmospheres in extreme conditions of temperature or pressure (Venus and Titan), and the exoplanet G1 851d under the considered three atmospheric conditions (see Table \ref{tab:table1}). 

\subsubsection{Atmospheres under conditions close to the ideal gas}

There are two circumstances in which a gas shows an ideal-gas behavior: (i) if its temperature is close to the Boyle temperature (the temperature value satisfying $B_2(T)=0$) \cite{tboyle}, and (ii) if it is a diluted gas \cite{mcquarriethermo}. The atmospheres of Earth and Mars are examples of these conditions, respectively.

\begin{figure}[ht!]
\centering
\includegraphics[scale=1]{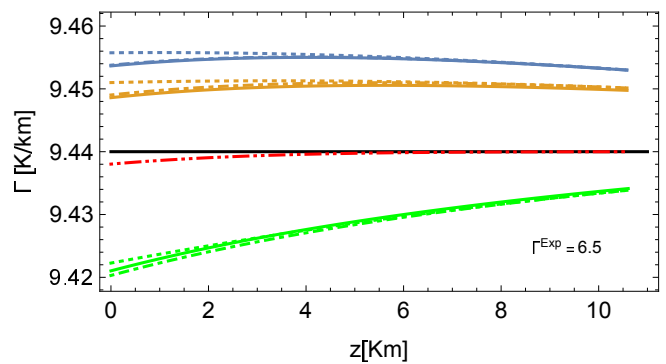}
\caption{Lapse rate for Earth. Lines for the ideal gas prediction with and without vibrations are denoted by black solid and red (gray) dot-dot-dashed lines, respectively. The considered virial expansions are denoted as follows: $B_2$ (dotted lines), $B_2$+vib (dash-dot lines), and $B_3$+vib (solid lines). Finally, colors represent the models as follows: van der Waals [blue (gray)], square-well [yellow (medium gray)], and hard-sphere [green (light gray)].}
\label{fig:lr-Tierra}
\end{figure}

In the case of Earth, neither the molecular interactions nor the molecular vibrations have a significant contribution to the DALR, as we show in Fig. \ref{fig:lr-Tierra}. On one hand, the vibration temperature of N$_2$ is $\approx \SI{3349}{\kelvin}$, which is much higher than the temperature on Earth's surface, $\SI{288}{\kelvin}$ (see Table. \ref{tab:table1}). The molecular vibrations contribution to the heat capacity is $0.00005$-$\SI{0.01}{\joule\per\kelvin}$ in the troposphere, which is negligible. On the other hand, Earth's temperature is close to the Boyle's temperature ($\approx 326.65$ and $\SI{327.51}{\kelvin}$ for van der Waals and square-well models, respectively). This means that attractive and repulsive interaction forces are almost balanced \cite{tboyle}. This makes the total molecular interactions negligible.

\begin{figure}[ht!]
\centering
\includegraphics[scale=1]{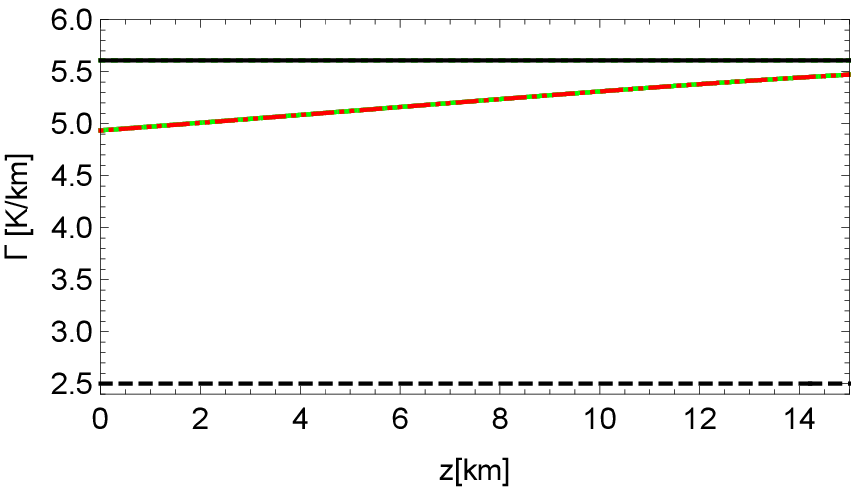}
\caption{Lapse rate for Mars. Lines and colors are the same as in Fig.~\ref{fig:lr-Tierra}. The observational value is represented by a black dashed line.}
\label{fig:lr-Marte}
\end{figure}

In Fig.~\ref{fig:lr-Marte} we show the DALR of Mars. Notice that the dominant contribution to the deviation of the lapse rate with respect to the ideal-gas prediction towards the observational value comes from the molecular vibrations. The reason for the negligible contribution from the virial coefficients is the low probability of observing molecular interactions since the atmosphere is diluted. This fact is consistent with the effects observed in the adiabatic curves (see Fig.~\ref{fig:adcurves} (d)).

\subsubsection{Atmospheres under extreme conditions of pressure or temperature} \label{sub extreme}

\begin{figure}[ht!]
\centering
\includegraphics[scale=1]{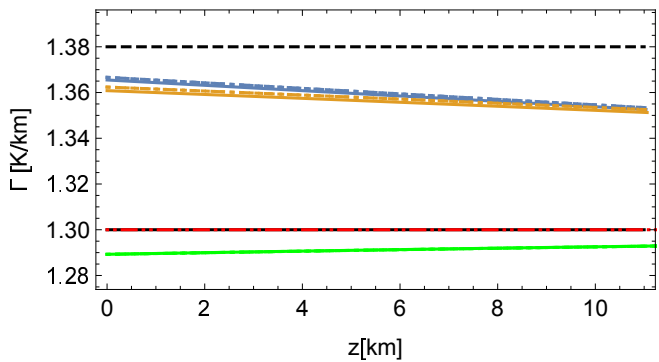}
\caption{Lapse rate for Titan. Lines and colors are the same as in Fig.~\ref{fig:lr-Tierra}. The observational value is represented by a black dashed line.}
\label{fig:lr-Titan}
\end{figure}

We call extreme conditions those that are close to conditions that allow phase transitions or to the vibrational temperatures, near the surface of the body. For example, comparing Titan's atmospheric conditions (see Table \ref{tab:table1}) with the boiling point of N$_2$ ($\approx \SI{77}{\kelvin}$ at $\SI{101}{\kilo\pascal}$) and its vibrational temperature ($\approx \SI{3349}{\kelvin}$), we expect that molecular interactions play a more important role than molecular vibrations in the resulting DALR. This is shown in Fig. \ref{fig:lr-Titan}. This fact is also observed in the adiabatic curves (see Fig. \ref{fig:adcurves} (b)). Notice that virial coefficients that also model attractive interactions (van der Waals and square-well) give values of DALR closer to the observational value. On the other hand, the hard sphere model, which represents a repulsive force only, gives a DALR that is worse than the ideal gas prediction.

\begin{figure}[ht!]
\centering
\includegraphics[scale=1]{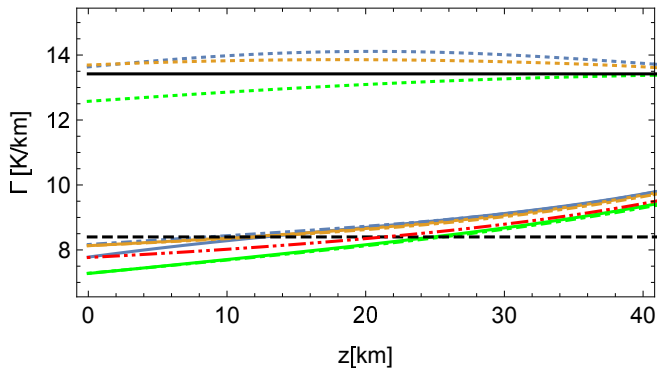}
\caption{Lapse rate for Venus. Lines and colors are the same as in Fig.~\ref{fig:lr-Tierra}. The observational value is represented by a black dashed line.}
\label{fig:lr-Venus}
\end{figure}

Venus is another planetary body under extreme conditions. In Fig. \ref{fig:lr-Venus} we show the value obtained for the DALRs. Notice that Venus' surface temperature is close to the first vibration temperature of CO$_2$ ($\approx \SI{960}{\kelvin}$). Therefore, the molecular vibrations have a more important effect on the DALR than the virial coefficients (we already observed this effect in the adiabatic curves in Fig. \ref{fig:adcurves} (c)). Remarkably, in those cases considering molecular vibrations, we obtain a DALR that is in a good agreement with the observed value.

\subsubsection{Exoplanet Gl 581d}

The exoplanet Gl 581d was discovered in the habitable zone of M-dwarf Gl 581 in 2007 \cite{Udry}.  According to Ref. \cite{HuDi}, it has around eight times the Earth’s mass. We emphasize that the range of atmospheric conditions considered in Table \ref{tab:table1} includes those allowing for the presence of liquid water on the planet surface \cite{HuDi}.

\begin{figure*}[ht!]
\centering
\includegraphics[scale=1]{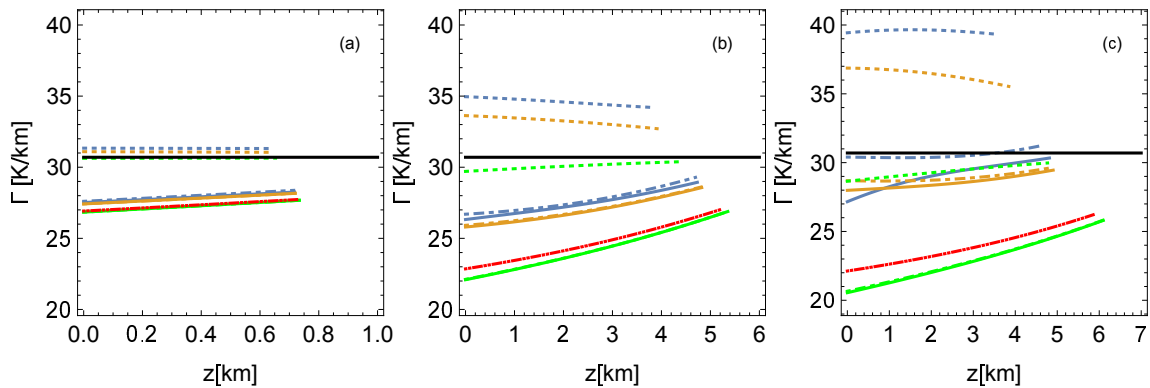}
\caption{Lapse rate for the exoplanet Gl 581d : A (a), B (b), and C (c) (see Table~\ref{tab:table1}). Lines and colors are the same as in Fig.~\ref{fig:lr-Tierra}.}
\label{fig:lr-exo}
\end{figure*}

This example allows us to illustrate, on the same astronomic object, how the atmospheric conditions dictate the contributions of molecular vibrations and interactions. In Fig. \ref{fig:lr-exo}, we show the DALR for Gl 581d under the atmospheric conditions A, B, and C in Table~\ref{tab:table1}.  For A, we observe that the dominant contribution to the DALR comes from the vibrations (as in the Mars case). On the other hand, in conditions B and C, which can be considered as extreme conditions, the molecular interactions also become relevant. Notice that in case C the contribution of vibrations and those of the van der Waals and square-well models are almost balanced, and then we get a DALR close to the ideal gas prediction. 

In this case, it is not possible to compare our results with any observational value because the real conditions for this exoplanet are unknown, but we can make comparisons with the ideal-gas prediction to observe the effect of the molecular interactions and vibrations. Additionally, one application of our results is the estimation of the height of the top of the troposphere. If we restrict the adiabatic curves to the region where no phase transitions are allowed, we obtain $\SI{1}{\kilo\metre}$, $\SI{5}{\kilo\metre}$, and $\SI{6}{\kilo\metre}$ for conditions A, B, and C, respectively. Notice that if the atmosphere contains traces of vapors (such as CO$_2$ or H$_2$O clouds) the use of the moist lapse rate is necessary (as in Earth). In this case, the estimated height of the top of the troposphere could increase.

\section{Discussion and conclusions} \label{sect: Dandc}

In this paper, we obtain a formula for the DALR that depends on the compressibility factor and the adiabatic curves. As our interest is to study the non-ideal behavior of atmospheric gases, we take into account the translation, rotation, and vibration of molecules as well as interactions between them. We consider the virial expansion for three models, namely, van der Waals, square-well, and hard-sphere. We analyze in detail the following cases: ideal gas, and virial expansion up the second order both with and without vibrations. We also consider third order contributions together with vibrational modes. We study the DALR for Earth, Mars, Venus, Titan, and the exoplanet Gl 581d under the previous circumstances. 

\begin{figure}[ht!]
\centering
\includegraphics[scale=1]{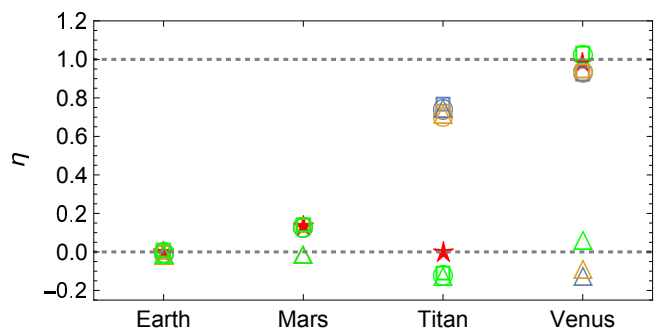}
\caption{Value of $\eta$ for the equations of state: I. G.+ \textrm{vib} [red (gray) stars], $B_2$ (triangles), $B_2$+\textrm{vib} (squares), $B_3$+\textrm{vib} (circles). The virial coefficient models are: van der Waals [blue (gray)], square-well [yellow (medium gray)], and hard-sphere [green (light gray)].}
\label{fig:eta}
\end{figure}

Notice that in all these cases the contribution of the molecular interactions to the DALR becomes negligible as the height increases. The reason is the decrease in the density, pressure, and temperature. In these conditions the ideal gas behavior is recovered. Regarding the observed value, $\Gamma^{\text{Obs}}$, we show it in the figures of Sec. \ref{results}, except for the exoplanet Gl 581d whose $\Gamma^{\text{Obs}}$ is unknown. To quantify how much our DALR approaches to $\Gamma^{\text{Obs}}$, we define the following auxiliary function:
\begin{equation}\label{eta}
\eta \coloneqq \frac{\Gamma^\text{IG}-\overline{\Gamma}}{\Gamma^{\text{IG}}-\Gamma^{\text{Obs}}}\,,
\end{equation}
where $\overline{\Gamma}$ denotes the quotient of $\Delta T$, the temperature difference between the surface and the top of the troposphere (calculated with $\Gamma$), over the corresponding $\Delta z$. In this way, we obtain a mean-type value for the DALR. In Fig.~\ref{fig:eta} we show the values of $\eta$ for the cases analyzed in Sec. \ref{sublr}.

The interpretation of $\eta$ is straightforward: if $\eta$ takes values around zero then $\overline{\Gamma}$ is close to $\Gamma^\text{IG}$. Earth and Mars are examples of this. We conclude that in these cases there are other contributions to the lapse rate that are more important than the molecular vibrations and interactions. For Earth, it is known that the atmosphere contains traces of vapors, and then we must use the moist lapse rate approach, which gives $6$-$\SI{7}{\kelvin\per\kilo\metre}$ in the middle layer of the troposphere \cite{Catling2015}, that is a better approximation to $\Gamma_\text{Obs}$. On Mars, we observe an improvement of $20\%$ by including molecular vibrations in the computation of the DALR, but it also has additional heating that comes from the absorption of solar radiation by suspended dust particles \cite{zurek_2017}.

On the other hand, $\eta\sim 1$ means that $\overline{\Gamma}$ is a very good approximation to $\Gamma_\text{Obs}$. This is the case for astronomical objects under extreme conditions for some models. In Titan, we have that the contributions of the interactions are more important than that of the vibrational ones, as we explain in Sec. \ref{sub extreme}. Conversely, for Venus, the vibrations are more important than molecular interactions (see Sec. \ref{sub extreme}).

Let us compare our approach to the results reported by Staley in Ref. \cite{Staley1970}, where the author analyzes Venus considering an atmosphere of $100\%$ CO$_2$ and using experimental data for the compressibility factor and heat capacity. Staley obtains a DALR of $\SI{7.8486}{\kelvin\per\kilo\metre}$ at $\SI{10100}{\kilo\pascal}$ and $\SI{750}{\kelvin}$. Under the same conditions, we obtain for the third order virial expansion and including vibrations the following values: $\SI{7.67024}{\kelvin\per\kilo\metre}$, $\SI{8.07806}{\kelvin\per\kilo\metre}$, and $\SI{7.18384}{\kelvin\per\kilo\metre}$, for van der Waals, square-well and hard-sphere models, respectively. The hard-sphere model gives the worst prediction. Notice that the difference between the first two models and the Staley prediction is less than $\pm\SI{0.24}{\kelvin\per\kilo\metre}$. It is worth mentioning that the square-well model gives the closest value to $\Gamma^{\textrm{Obs}}$.

Finally, as we state in Sec. \ref{sect: 2}, our approach can be applied for the compressibility factor of other equations of state, not only in the virial expansion, to incorporate the molecular interactions. We intend to analyze this in a future work. Furthermore, a feasible extension to this paper is to take into account that atmospheres are composed of several gases (mixed gases). This modifies the molecular mass, the heat capacity, and the virial coefficients $B_k^\text{mix}$, that now take into account all possible interactions into the mixed gases \cite{Dy2003mix}.  A step forward in the understanding of the lapse rate is to include the molecular vibrations and interactions in other approaches to the computation of the lapse rate, for instance, the moist lapse rate. However, to be applied, these generalizations could face the problem of data availability.

\begin{acknowledgments}

Bogar Díaz and J. E. Ramírez are supported by a Consejo Nacional de Ciencia y Tecnolog\'ia postdoctoral fellowship (Grants
No. 371778 and No. 289198). The numerical calculations and plots were carried out using Wolfram MATHEMATICA 12. We would like to thank Miguel Ángel García-Ariza, J. Fernando Barbero G., Carlos Pajares, and M. Iván Martínez for fruitful discussions and valuable comments.
\end{acknowledgments}


\bibliography{main}

\begin{thebibliography}{37}%
\makeatletter
\providecommand \@ifxundefined [1]{%
 \@ifx{#1\undefined}
}%
\providecommand \@ifnum [1]{%
 \ifnum #1\expandafter \@firstoftwo
 \else \expandafter \@secondoftwo
 \fi
}%
\providecommand \@ifx [1]{%
 \ifx #1\expandafter \@firstoftwo
 \else \expandafter \@secondoftwo
 \fi
}%
\providecommand \natexlab [1]{#1}%
\providecommand \enquote  [1]{``#1''}%
\providecommand \bibnamefont  [1]{#1}%
\providecommand \bibfnamefont [1]{#1}%
\providecommand \citenamefont [1]{#1}%
\providecommand \href@noop [0]{\@secondoftwo}%
\providecommand \href [0]{\begingroup \@sanitize@url \@href}%
\providecommand \@href[1]{\@@startlink{#1}\@@href}%
\providecommand \@@href[1]{\endgroup#1\@@endlink}%
\providecommand \@sanitize@url [0]{\catcode `\\12\catcode `\$12\catcode
  `\&12\catcode `\#12\catcode `\^12\catcode `\_12\catcode `\%12\relax}%
\providecommand \@@startlink[1]{}%
\providecommand \@@endlink[0]{}%
\providecommand \url  [0]{\begingroup\@sanitize@url \@url }%
\providecommand \@url [1]{\endgroup\@href {#1}{\urlprefix }}%
\providecommand \urlprefix  [0]{URL }%
\providecommand \Eprint [0]{\href }%
\providecommand \doibase [0]{https://doi.org/}%
\providecommand \selectlanguage [0]{\@gobble}%
\providecommand \bibinfo  [0]{\@secondoftwo}%
\providecommand \bibfield  [0]{\@secondoftwo}%
\providecommand \translation [1]{[#1]}%
\providecommand \BibitemOpen [0]{}%
\providecommand \bibitemStop [0]{}%
\providecommand \bibitemNoStop [0]{.\EOS\space}%
\providecommand \EOS [0]{\spacefactor3000\relax}%
\providecommand \BibitemShut  [1]{\csname bibitem#1\endcsname}%
\let\auto@bib@innerbib\@empty
\bibitem [{\citenamefont {Kasprzak}(1990)}]{Kasprzak1990}%
  \BibitemOpen
  \bibfield  {author} {\bibinfo {author} {\bibfnamefont {W.~T.}\ \bibnamefont
  {Kasprzak}},\ }\href@noop {} {\emph {\bibinfo {title} {The {P}ioneer {V}enus
  {O}rbiter: 11 years of {D}ata. {A} laboratory for atmospheres seminar
  talk}}},\ \bibinfo {type} {Tech. Rep.}\ \bibinfo {number} {NASA-TM-100761}\
  (\bibinfo  {institution} {NASA Goddard Space Flight Center},\ \bibinfo
  {address} {Greenbelt, MD, USA},\ \bibinfo {year} {1990})\BibitemShut
  {NoStop}%
\bibitem [{\citenamefont {Lindal}\ \emph {et~al.}(1983)\citenamefont {Lindal},
  \citenamefont {Wood}, \citenamefont {Hotz}, \citenamefont {Sweetnam},
  \citenamefont {Eshleman},\ and\ \citenamefont {Tyler}}]{Lindal1983}%
  \BibitemOpen
  \bibfield  {author} {\bibinfo {author} {\bibfnamefont {G.~F.}\ \bibnamefont
  {Lindal}}, \bibinfo {author} {\bibfnamefont {G.~E.}\ \bibnamefont {Wood}},
  \bibinfo {author} {\bibfnamefont {H.~B.}\ \bibnamefont {Hotz}}, \bibinfo
  {author} {\bibfnamefont {D.~N.}\ \bibnamefont {Sweetnam}}, \bibinfo {author}
  {\bibfnamefont {V.~R.}\ \bibnamefont {Eshleman}},\ and\ \bibinfo {author}
  {\bibfnamefont {G.~L.}\ \bibnamefont {Tyler}},\ }\bibfield  {title} {\bibinfo
  {title} {The atmosphere of {T}itan: {A}n analysis of the {V}oyager 1 radio
  occultation measurements},\ }\href
  {https://doi.org/https://doi.org/10.1016/0019-1035(83)90155-0} {\bibfield
  {journal} {\bibinfo  {journal} {Icarus}\ }\textbf {\bibinfo {volume} {53}},\
  \bibinfo {pages} {348 } (\bibinfo {year} {1983})}\BibitemShut {NoStop}%
\bibitem [{\citenamefont {Mokhov}\ and\ \citenamefont
  {Akperov}(2006)}]{Mokhov2006}%
  \BibitemOpen
  \bibfield  {author} {\bibinfo {author} {\bibfnamefont {I.~I.}\ \bibnamefont
  {Mokhov}}\ and\ \bibinfo {author} {\bibfnamefont {M.~G.}\ \bibnamefont
  {Akperov}},\ }\bibfield  {title} {\bibinfo {title} {Tropospheric lapse rate
  and its relation to surface temperature from reanalysis data},\ }\href
  {https://doi.org/10.1134/S0001433806040037} {\bibfield  {journal} {\bibinfo
  {journal} {Izv. Atmos. Ocean. Phys.}\ }\textbf {\bibinfo {volume} {42}},\
  \bibinfo {pages} {430} (\bibinfo {year} {2006})}\BibitemShut {NoStop}%
\bibitem [{\citenamefont {Zurek}(2017)}]{zurek_2017}%
  \BibitemOpen
  \bibfield  {author} {\bibinfo {author} {\bibfnamefont {R.~W.}\ \bibnamefont
  {Zurek}},\ }\bibinfo {title} {Understanding {M}ars and {I}ts {A}tmosphere},\
  in\ \href {https://doi.org/10.1017/9781139060172.002} {\emph {\bibinfo
  {booktitle} {{The Atmosphere and Climate of Mars}}}},\ \bibinfo {series and
  number} {Cambridge Planetary Science},\ \bibinfo {editor} {edited by\
  \bibinfo {editor} {\bibfnamefont {R.~M.}\ \bibnamefont {Haberle}}, \bibinfo
  {editor} {\bibfnamefont {R.~T.}\ \bibnamefont {Clancy}}, \bibinfo {editor}
  {\bibfnamefont {F.}~\bibnamefont {Forget}}, \bibinfo {editor} {\bibfnamefont
  {M.~D.}\ \bibnamefont {Smith}},\ and\ \bibinfo {editor} {\bibfnamefont
  {R.~W.}\ \bibnamefont {Zurek}}}\ (\bibinfo  {publisher} {Cambridge University
  Press},\ \bibinfo {year} {2017})\ p.\ \bibinfo {pages} {3–19}\BibitemShut
  {NoStop}%
\bibitem [{\citenamefont {Grady}(2003)}]{astrobio}%
  \BibitemOpen
  \bibfield  {author} {\bibinfo {author} {\bibfnamefont {M.~M.}\ \bibnamefont
  {Grady}},\ }\bibfield  {title} {\bibinfo {title} {Feature},\ }\href
  {https://doi.org/10.1046/j.1365-2451.2003.00403.x} {\bibfield  {journal}
  {\bibinfo  {journal} {Geology Today}\ }\textbf {\bibinfo {volume} {19}},\
  \bibinfo {pages} {99} (\bibinfo {year} {2003})}\BibitemShut {NoStop}%
\bibitem [{\citenamefont {{Y. Hu}}\ and\ \citenamefont {{ F.
  Ding}}(2011)}]{HuDi}%
  \BibitemOpen
  \bibfield  {author} {\bibinfo {author} {\bibnamefont {{Y. Hu}}}\ and\
  \bibinfo {author} {\bibnamefont {{ F. Ding}}},\ }\bibfield  {title} {\bibinfo
  {title} {Radiative constraints on the habitability of exoplanets
  {G}liese~581c and {G}liese~581d},\ }\href
  {https://doi.org/10.1051/0004-6361/201014880} {\bibfield  {journal} {\bibinfo
   {journal} {A\&A}\ }\textbf {\bibinfo {volume} {526}},\ \bibinfo {pages}
  {A135} (\bibinfo {year} {2011})}\BibitemShut {NoStop}%
\bibitem [{\citenamefont {Catling}(2015)}]{Catling2015}%
  \BibitemOpen
  \bibfield  {author} {\bibinfo {author} {\bibfnamefont {D.~C.}\ \bibnamefont
  {Catling}},\ }\bibfield  {title} {\bibinfo {title} {{Planetary
  Atmospheres}},\ }in\ \href
  {https://doi.org/https://doi.org/10.1016/B978-0-444-53802-4.00185-8} {\emph
  {\bibinfo {booktitle} {Treatise on Geophysics (Second Edition)}}},\ \bibinfo
  {editor} {edited by\ \bibinfo {editor} {\bibfnamefont {G.}~\bibnamefont
  {Schubert}}}\ (\bibinfo  {publisher} {Elsevier},\ \bibinfo {address}
  {Oxford},\ \bibinfo {year} {2015})\ \bibinfo {edition} {second edition}\
  ed.,\ pp.\ \bibinfo {pages} {429 -- 472}\BibitemShut {NoStop}%
\bibitem [{\citenamefont {Vallero}(2014)}]{Vallero2014}%
  \BibitemOpen
  \bibfield  {author} {\bibinfo {author} {\bibfnamefont {D.}~\bibnamefont
  {Vallero}},\ }\bibfield  {title} {\bibinfo {title} {{The Physics of the
  Atmosphere}},\ }in\ \href
  {https://doi.org/https://doi.org/10.1016/B978-0-12-401733-7.00002-5} {\emph
  {\bibinfo {booktitle} {Fundamentals of Air Pollution (Fifth Edition)}}},\
  \bibinfo {editor} {edited by\ \bibinfo {editor} {\bibfnamefont
  {D.}~\bibnamefont {Vallero}}}\ (\bibinfo  {publisher} {Academic Press},\
  \bibinfo {address} {Boston},\ \bibinfo {year} {2014})\ \bibinfo {edition}
  {fifth edition}\ ed.,\ pp.\ \bibinfo {pages} {23 -- 42}\BibitemShut {NoStop}%
\bibitem [{\citenamefont {Staley}(1970)}]{Staley1970}%
  \BibitemOpen
  \bibfield  {author} {\bibinfo {author} {\bibfnamefont {D.~O.}\ \bibnamefont
  {Staley}},\ }\bibfield  {title} {\bibinfo {title} {The {A}diabatic {L}apse
  {R}ate in the {V}enus {A}tmosphere},\ }\href@noop {} {\bibfield  {journal}
  {\bibinfo  {journal} {J. Atmos. Sci.}\ }\textbf {\bibinfo {volume} {27}}
  (\bibinfo {year} {1970})}\BibitemShut {NoStop}%
\bibitem [{\citenamefont {Alvarez~Navarro}\ \emph {et~al.}(2019)\citenamefont
  {Alvarez~Navarro}, \citenamefont {Díaz}, \citenamefont {García-Ariza},\
  and\ \citenamefont {Ramírez}}]{JHON}%
  \BibitemOpen
  \bibfield  {author} {\bibinfo {author} {\bibfnamefont {E.}~\bibnamefont
  {Alvarez~Navarro}}, \bibinfo {author} {\bibfnamefont {B.}~\bibnamefont
  {Díaz}}, \bibinfo {author} {\bibfnamefont {M.~A.}\ \bibnamefont
  {García-Ariza}},\ and\ \bibinfo {author} {\bibfnamefont {J.~E.}\
  \bibnamefont {Ramírez}},\ }\bibfield  {title} {\bibinfo {title} {Effects of
  the second virial coefficient on the adiabatic lapse rate of dry
  atmospheres},\ }\href {https://doi.org/10.1140/epjp/i2019-12826-4} {\bibfield
   {journal} {\bibinfo  {journal} {Eur. Phys. J. Plus}\ }\textbf {\bibinfo
  {volume} {134}},\ \bibinfo {pages} {458} (\bibinfo {year}
  {2019})}\BibitemShut {NoStop}%
\bibitem [{\citenamefont {Schroeder}(2000)}]{schroeder2000}%
  \BibitemOpen
  \bibfield  {author} {\bibinfo {author} {\bibfnamefont {D.}~\bibnamefont
  {Schroeder}},\ }\href@noop {} {\emph {\bibinfo {title} {{Introduction to
  Thermal Physics}}}}\ (\bibinfo  {publisher} {Addison-Wesley Longman,
  Incorporated},\ \bibinfo {year} {2000})\BibitemShut {NoStop}%
\bibitem [{Note1()}]{Note1}%
  \BibitemOpen
  \bibinfo {note} {Using Newton's law of universal gravitation the acceleration
  due to gravity as a function of the high $z$ can be written as $g(z)= g /
  \left (1+ z/r \right )^2$, where $r$ denotes the radio of the astronomic
  object.}\BibitemShut {Stop}%
\bibitem [{\citenamefont {McQuarrie}\ and\ \citenamefont
  {Simon}(1999)}]{mcquarriethermo}%
  \BibitemOpen
  \bibfield  {author} {\bibinfo {author} {\bibfnamefont {D.}~\bibnamefont
  {McQuarrie}}\ and\ \bibinfo {author} {\bibfnamefont {J.}~\bibnamefont
  {Simon}},\ }\href {https://books.google.es/books?id=TqAIJ27C2isC} {\emph
  {\bibinfo {title} {Molecular {T}hermodynamics}}}\ (\bibinfo  {publisher}
  {University Science Books},\ \bibinfo {year} {1999})\BibitemShut {NoStop}%
\bibitem [{\citenamefont {Lemmon}\ \emph {et~al.}(2018)\citenamefont {Lemmon},
  , \citenamefont {Bell}, \citenamefont {Huber},\ and\ \citenamefont
  {McLinden}}]{dataZ}%
  \BibitemOpen
  \bibfield  {author} {\bibinfo {author} {\bibfnamefont {E.~W.}\ \bibnamefont
  {Lemmon}}, , \bibinfo {author} {\bibfnamefont {I.~H.}\ \bibnamefont {Bell}},
  \bibinfo {author} {\bibfnamefont {M.~L.}\ \bibnamefont {Huber}},\ and\
  \bibinfo {author} {\bibfnamefont {M.~O.}\ \bibnamefont {McLinden}},\ }\href
  {https://doi.org/http://dx.doi.org/10.18434/T4JS3C} {\bibinfo {title} {NIST
  Standard Reference Database 23: Reference Fluid Thermodynamic and Transport
  Properties-REFPROP, Version 10.0, National Institute of Standards and
  Technology}} (\bibinfo {year} {2018})\BibitemShut {NoStop}%
\bibitem [{\citenamefont {Cengel}\ \emph {et~al.}(2018)\citenamefont {Cengel},
  \citenamefont {Boles},\ and\ \citenamefont {Kanoglu}}]{cengel}%
  \BibitemOpen
  \bibfield  {author} {\bibinfo {author} {\bibfnamefont {Y.~A.}\ \bibnamefont
  {Cengel}}, \bibinfo {author} {\bibfnamefont {M.}~\bibnamefont {Boles}},\ and\
  \bibinfo {author} {\bibfnamefont {M.}~\bibnamefont {Kanoglu}},\ }\href@noop
  {} {\emph {\bibinfo {title} {{Thermodynamics: An Engineering Approach}}}}\
  (\bibinfo  {publisher} {McGraw-Hill Higher Education},\ \bibinfo {year}
  {2018})\BibitemShut {NoStop}%
\bibitem [{\citenamefont {McQuarrie}(2000)}]{mcquarrie}%
  \BibitemOpen
  \bibfield  {author} {\bibinfo {author} {\bibfnamefont {D.}~\bibnamefont
  {McQuarrie}},\ }\href@noop {} {\emph {\bibinfo {title} {{Statistical
  Mechanics}}}}\ (\bibinfo  {publisher} {University Science Books},\ \bibinfo
  {year} {2000})\BibitemShut {NoStop}%
\bibitem [{\citenamefont {Mayer}(1958)}]{Mayer1958}%
  \BibitemOpen
  \bibfield  {author} {\bibinfo {author} {\bibfnamefont {J.~E.}\ \bibnamefont
  {Mayer}},\ }\bibinfo {title} {Theory of real gases},\ in\ \href
  {https://doi.org/10.1007/978-3-642-45892-7_2} {\emph {\bibinfo {booktitle}
  {{Thermodynamik der Gase / Thermodynamics of Gases}}}},\ \bibinfo {editor}
  {edited by\ \bibinfo {editor} {\bibfnamefont {S.}~\bibnamefont
  {Fl{\"u}gge}}}\ (\bibinfo  {publisher} {Springer Berlin Heidelberg},\
  \bibinfo {address} {Berlin, Heidelberg},\ \bibinfo {year} {1958})\ pp.\
  \bibinfo {pages} {73--204}\BibitemShut {NoStop}%
\bibitem [{\citenamefont {Ushcats}(2012)}]{mayerVirial}%
  \BibitemOpen
  \bibfield  {author} {\bibinfo {author} {\bibfnamefont {M.~V.}\ \bibnamefont
  {Ushcats}},\ }\bibfield  {title} {\bibinfo {title} {{Equation of State Beyond
  the Radius of Convergence of the Virial Expansion}},\ }\href
  {https://doi.org/10.1103/PhysRevLett.109.040601} {\bibfield  {journal}
  {\bibinfo  {journal} {Phys. Rev. Lett.}\ }\textbf {\bibinfo {volume} {109}},\
  \bibinfo {pages} {040601} (\bibinfo {year} {2012})}\BibitemShut {NoStop}%
\bibitem [{\citenamefont {Ushcats}\ \emph {et~al.}(2017)\citenamefont
  {Ushcats}, \citenamefont {Bulavin}, \citenamefont {Sysoev},\ and\
  \citenamefont {Ushcats}}]{ushcats1}%
  \BibitemOpen
  \bibfield  {author} {\bibinfo {author} {\bibfnamefont {M.~V.}\ \bibnamefont
  {Ushcats}}, \bibinfo {author} {\bibfnamefont {L.~A.}\ \bibnamefont
  {Bulavin}}, \bibinfo {author} {\bibfnamefont {V.~M.}\ \bibnamefont
  {Sysoev}},\ and\ \bibinfo {author} {\bibfnamefont {S.~Y.}\ \bibnamefont
  {Ushcats}},\ }\bibfield  {title} {\bibinfo {title} {Divergence of activity
  expansions: Is it actually a problem?},\ }\href
  {https://doi.org/10.1103/PhysRevE.96.062115} {\bibfield  {journal} {\bibinfo
  {journal} {Phys. Rev. E}\ }\textbf {\bibinfo {volume} {96}},\ \bibinfo
  {pages} {062115} (\bibinfo {year} {2017})}\BibitemShut {NoStop}%
\bibitem [{\citenamefont {Ushcats}\ \emph
  {et~al.}(2018{\natexlab{a}})\citenamefont {Ushcats}, \citenamefont
  {Bulavin},\ and\ \citenamefont {Ushcats}}]{ushcats2}%
  \BibitemOpen
  \bibfield  {author} {\bibinfo {author} {\bibfnamefont {M.~V.}\ \bibnamefont
  {Ushcats}}, \bibinfo {author} {\bibfnamefont {L.~A.}\ \bibnamefont
  {Bulavin}},\ and\ \bibinfo {author} {\bibfnamefont {S.~Y.}\ \bibnamefont
  {Ushcats}},\ }\bibfield  {title} {\bibinfo {title} {Evidence for a
  first-order phase transition at the divergence region of activity
  expansions},\ }\href {https://doi.org/10.1103/PhysRevE.98.042127} {\bibfield
  {journal} {\bibinfo  {journal} {Phys. Rev. E}\ }\textbf {\bibinfo {volume}
  {98}},\ \bibinfo {pages} {042127} (\bibinfo {year}
  {2018}{\natexlab{a}})}\BibitemShut {NoStop}%
\bibitem [{\citenamefont {Ushcats}\ and\ \citenamefont
  {Bulavin}(2020)}]{ushcats3}%
  \BibitemOpen
  \bibfield  {author} {\bibinfo {author} {\bibfnamefont {M.~V.}\ \bibnamefont
  {Ushcats}}\ and\ \bibinfo {author} {\bibfnamefont {L.~A.}\ \bibnamefont
  {Bulavin}},\ }\bibfield  {title} {\bibinfo {title} {Construction of
  subcritical isotherms for model and real gases on the basis of mayer's
  cluster expansion},\ }\href {https://doi.org/10.1103/PhysRevE.101.062128}
  {\bibfield  {journal} {\bibinfo  {journal} {Phys. Rev. E}\ }\textbf {\bibinfo
  {volume} {101}},\ \bibinfo {pages} {062128} (\bibinfo {year}
  {2020})}\BibitemShut {NoStop}%
\bibitem [{\citenamefont {Mayer}\ and\ \citenamefont
  {Mayer}(1940)}]{mayerbook}%
  \BibitemOpen
  \bibfield  {author} {\bibinfo {author} {\bibfnamefont {J.}~\bibnamefont
  {Mayer}}\ and\ \bibinfo {author} {\bibfnamefont {M.}~\bibnamefont {Mayer}},\
  }\href@noop {} {\emph {\bibinfo {title} {{Statistical Mechanics}}}}\
  (\bibinfo  {publisher} {J. Wiley \& Sons, Incorporated},\ \bibinfo {year}
  {1940})\BibitemShut {NoStop}%
\bibitem [{\citenamefont {Ushcats}\ \emph
  {et~al.}(2018{\natexlab{b}})\citenamefont {Ushcats}, \citenamefont {Ushcats},
  \citenamefont {Bulavin},\ and\ \citenamefont {Sysoev}}]{ushcats4}%
  \BibitemOpen
  \bibfield  {author} {\bibinfo {author} {\bibfnamefont {M.~V.}\ \bibnamefont
  {Ushcats}}, \bibinfo {author} {\bibfnamefont {S.~Y.}\ \bibnamefont
  {Ushcats}}, \bibinfo {author} {\bibfnamefont {L.~A.}\ \bibnamefont
  {Bulavin}},\ and\ \bibinfo {author} {\bibfnamefont {V.~M.}\ \bibnamefont
  {Sysoev}},\ }\bibfield  {title} {\bibinfo {title} {Equation of state for all
  regimes of a fluid: From gas to liquid},\ }\href
  {https://doi.org/10.1103/PhysRevE.98.032135} {\bibfield  {journal} {\bibinfo
  {journal} {Phys. Rev. E}\ }\textbf {\bibinfo {volume} {98}},\ \bibinfo
  {pages} {032135} (\bibinfo {year} {2018}{\natexlab{b}})}\BibitemShut
  {NoStop}%
\bibitem [{Note2()}]{Note2}%
  \BibitemOpen
  \bibinfo {note} {The Lambert function $W$ is defined by $W(z) \protect
  \qopname \relax o{exp}\left ( W(z) \right ) = z$ and it cannot be expressed
  in terms of elementary functions.}\BibitemShut {Stop}%
\bibitem [{\citenamefont {Ree}\ and\ \citenamefont {Hoover}(1964)}]{vcHS}%
  \BibitemOpen
  \bibfield  {author} {\bibinfo {author} {\bibfnamefont {F.~H.}\ \bibnamefont
  {Ree}}\ and\ \bibinfo {author} {\bibfnamefont {W.~G.}\ \bibnamefont
  {Hoover}},\ }\bibfield  {title} {\bibinfo {title} {Fifth and {S}ixth {V}irial
  {C}oefficients for {H}ard {S}pheres and {H}ard {D}isks},\ }\href
  {https://doi.org/10.1063/1.1725286} {\bibfield  {journal} {\bibinfo
  {journal} {J. Chem. Phys.}\ }\textbf {\bibinfo {volume} {40}},\ \bibinfo
  {pages} {939} (\bibinfo {year} {1964})}\BibitemShut {NoStop}%
\bibitem [{\citenamefont {Ree}\ and\ \citenamefont {Hoover}(1967)}]{vcHS2}%
  \BibitemOpen
  \bibfield  {author} {\bibinfo {author} {\bibfnamefont {F.~H.}\ \bibnamefont
  {Ree}}\ and\ \bibinfo {author} {\bibfnamefont {W.~G.}\ \bibnamefont
  {Hoover}},\ }\bibfield  {title} {\bibinfo {title} {Seventh {V}irial
  {C}oefficients for {H}ard {S}pheres and {H}ard {D}isks},\ }\href
  {https://doi.org/10.1063/1.1840521} {\bibfield  {journal} {\bibinfo
  {journal} {J. Chem. Phys.}\ }\textbf {\bibinfo {volume} {46}},\ \bibinfo
  {pages} {4181} (\bibinfo {year} {1967})}\BibitemShut {NoStop}%
\bibitem [{\citenamefont {Kihara}(1953)}]{vcSW}%
  \BibitemOpen
  \bibfield  {author} {\bibinfo {author} {\bibfnamefont {T.}~\bibnamefont
  {Kihara}},\ }\bibfield  {title} {\bibinfo {title} {Virial {C}oefficients and
  {M}odels of {M}olecules in {G}ases},\ }\href
  {https://doi.org/10.1103/RevModPhys.25.831} {\bibfield  {journal} {\bibinfo
  {journal} {Rev. Mod. Phys.}\ }\textbf {\bibinfo {volume} {25}},\ \bibinfo
  {pages} {831} (\bibinfo {year} {1953})}\BibitemShut {NoStop}%
\bibitem [{\citenamefont {Hussein}\ and\ \citenamefont {Ahmed}(1991)}]{vcSW2}%
  \BibitemOpen
  \bibfield  {author} {\bibinfo {author} {\bibfnamefont {N.~A.~R.}\
  \bibnamefont {Hussein}}\ and\ \bibinfo {author} {\bibfnamefont {S.~M.}\
  \bibnamefont {Ahmed}},\ }\bibfield  {title} {\bibinfo {title} {Virial
  coefficients for the square-well potential},\ }\href
  {https://doi.org/10.1088/0305-4470/24/1/035} {\bibfield  {journal} {\bibinfo
  {journal} {J. Phys. A-Math. Gen.}\ }\textbf {\bibinfo {volume} {24}},\
  \bibinfo {pages} {289} (\bibinfo {year} {1991})}\BibitemShut {NoStop}%
\bibitem [{\citenamefont {Dymond}\ \emph {et~al.}(2002)\citenamefont {Dymond},
  \citenamefont {Marsh},\ and\ \citenamefont {Wilhoit}}]{dymond2002virial}%
  \BibitemOpen
  \bibfield  {author} {\bibinfo {author} {\bibfnamefont {J.~D.}\ \bibnamefont
  {Dymond}}, \bibinfo {author} {\bibfnamefont {K.~N.}\ \bibnamefont {Marsh}},\
  and\ \bibinfo {author} {\bibfnamefont {R.~C.}\ \bibnamefont {Wilhoit}},\
  }\href@noop {} {\emph {\bibinfo {title} {Virial coefficients of pure
  gases}}},\ Vol.~\bibinfo {volume} {21}\ (\bibinfo  {publisher} {Springer
  Landord-Bornstein},\ \bibinfo {year} {2002})\BibitemShut {NoStop}%
\bibitem [{\citenamefont {{R. D. Wordsworth}}\ \emph
  {et~al.}(2010)\citenamefont {{R. D. Wordsworth}}, \citenamefont {{F.
  Forget}}, \citenamefont {{F. Selsis}}, \citenamefont {{J.-B. Madeleine}},
  \citenamefont {{E. Millour}},\ and\ \citenamefont {{V. Eymet}}}]{Words}%
  \BibitemOpen
  \bibfield  {author} {\bibinfo {author} {\bibnamefont {{R. D. Wordsworth}}},
  \bibinfo {author} {\bibnamefont {{F. Forget}}}, \bibinfo {author}
  {\bibnamefont {{F. Selsis}}}, \bibinfo {author} {\bibnamefont {{J.-B.
  Madeleine}}}, \bibinfo {author} {\bibnamefont {{E. Millour}}},\ and\ \bibinfo
  {author} {\bibnamefont {{V. Eymet}}},\ }\bibfield  {title} {\bibinfo {title}
  {{Is {G}liese 581d habitable? Some constraints from radiative-convective
  climate modeling}},\ }\href {https://doi.org/10.1051/0004-6361/201015053}
  {\bibfield  {journal} {\bibinfo  {journal} {A\&A}\ }\textbf {\bibinfo
  {volume} {522}},\ \bibinfo {pages} {A22} (\bibinfo {year}
  {2010})}\BibitemShut {NoStop}%
\bibitem [{\citenamefont {{P. von Paris}}\ \emph {et~al.}(2010)\citenamefont
  {{P. von Paris}}, \citenamefont {{S. Gebauer}}, \citenamefont {{M. Godolt}},
  \citenamefont {{J. L. Grenfell}}, \citenamefont {{P. Hedelt}}, \citenamefont
  {{D. Kitzmann}}, \citenamefont {{A. B. C. Patzer}}, \citenamefont {{H.
  Rauer}},\ and\ \citenamefont {{B. Stracke}}}]{von}%
  \BibitemOpen
  \bibfield  {author} {\bibinfo {author} {\bibnamefont {{P. von Paris}}},
  \bibinfo {author} {\bibnamefont {{S. Gebauer}}}, \bibinfo {author}
  {\bibnamefont {{M. Godolt}}}, \bibinfo {author} {\bibnamefont {{J. L.
  Grenfell}}}, \bibinfo {author} {\bibnamefont {{P. Hedelt}}}, \bibinfo
  {author} {\bibnamefont {{D. Kitzmann}}}, \bibinfo {author} {\bibnamefont {{A.
  B. C. Patzer}}}, \bibinfo {author} {\bibnamefont {{H. Rauer}}},\ and\
  \bibinfo {author} {\bibnamefont {{B. Stracke}}},\ }\bibfield  {title}
  {\bibinfo {title} {The extrasolar planet {G}liese 581d: a potentially
  habitable planet?},\ }\href {https://doi.org/10.1051/0004-6361/201015329}
  {\bibfield  {journal} {\bibinfo  {journal} {A\&A}\ }\textbf {\bibinfo
  {volume} {522}},\ \bibinfo {pages} {A23} (\bibinfo {year}
  {2010})}\BibitemShut {NoStop}%
\bibitem [{\citenamefont {van~den Bekerom}\ \emph {et~al.}(2018)\citenamefont
  {van~den Bekerom}, \citenamefont {Linares}, \citenamefont {van Veldhuizen},
  \citenamefont {Nijdam}, \citenamefont {van~de Sanden},\ and\ \citenamefont
  {van Rooij}}]{dataCO2}%
  \BibitemOpen
  \bibfield  {author} {\bibinfo {author} {\bibfnamefont {D.~C.~M.}\
  \bibnamefont {van~den Bekerom}}, \bibinfo {author} {\bibfnamefont {J.~M.~P.}\
  \bibnamefont {Linares}}, \bibinfo {author} {\bibfnamefont {E.~M.}\
  \bibnamefont {van Veldhuizen}}, \bibinfo {author} {\bibfnamefont
  {S.}~\bibnamefont {Nijdam}}, \bibinfo {author} {\bibfnamefont {M.~C.~M.}\
  \bibnamefont {van~de Sanden}},\ and\ \bibinfo {author} {\bibfnamefont
  {G.~J.}\ \bibnamefont {van Rooij}},\ }\bibfield  {title} {\bibinfo {title}
  {{How the alternating degeneracy in rotational Raman spectra of {CO}$_2$ and
  {C}$_2${H}$_2$ reveals the vibrational temperature}},\ }\href
  {https://doi.org/10.1364/AO.57.005694} {\bibfield  {journal} {\bibinfo
  {journal} {Appl. Opt.}\ }\textbf {\bibinfo {volume} {57}},\ \bibinfo {pages}
  {5694} (\bibinfo {year} {2018})}\BibitemShut {NoStop}%
\bibitem [{\citenamefont {Petrov}\ \emph {et~al.}(2018)\citenamefont {Petrov},
  \citenamefont {Matrosov}, \citenamefont {Sedinkin},\ and\ \citenamefont
  {Zaripov}}]{dataN2}%
  \BibitemOpen
  \bibfield  {author} {\bibinfo {author} {\bibfnamefont {D.~V.}\ \bibnamefont
  {Petrov}}, \bibinfo {author} {\bibfnamefont {I.~I.}\ \bibnamefont
  {Matrosov}}, \bibinfo {author} {\bibfnamefont {D.~O.}\ \bibnamefont
  {Sedinkin}},\ and\ \bibinfo {author} {\bibfnamefont {A.~R.}\ \bibnamefont
  {Zaripov}},\ }\bibfield  {title} {\bibinfo {title} {{Raman Spectra of
  Nitrogen, Carbon Dioxide, and Hydrogen in a Methane Environment}},\
  }\href@noop {} {\bibfield  {journal} {\bibinfo  {journal} {Opt. Spectrosc.}\
  }\textbf {\bibinfo {volume} {124}},\ \bibinfo {pages} {8} (\bibinfo {year}
  {2018})}\BibitemShut {NoStop}%
\bibitem [{\citenamefont {Span}\ and\ \citenamefont {Wagner}(1996)}]{Span}%
  \BibitemOpen
  \bibfield  {author} {\bibinfo {author} {\bibfnamefont {R.}~\bibnamefont
  {Span}}\ and\ \bibinfo {author} {\bibfnamefont {W.}~\bibnamefont {Wagner}},\
  }\bibfield  {title} {\bibinfo {title} {{A New Equation of State for Carbon
  Dioxide Covering the Fluid Region from the Triple-Point Temperature to $1100$
  {K} at Pressures up to $800$ {MP}a}},\ }\href
  {https://doi.org/10.1063/1.555991} {\bibfield  {journal} {\bibinfo  {journal}
  {J. Phys. Chem. Ref. Data}\ }\textbf {\bibinfo {volume} {25}},\ \bibinfo
  {pages} {1509} (\bibinfo {year} {1996})}\BibitemShut {NoStop}%
\bibitem [{\citenamefont {Coccia}\ \emph {et~al.}(2019)\citenamefont {Coccia},
  \citenamefont {Nicola}, \citenamefont {Tomassetti}, \citenamefont
  {Pierantozzi},\ and\ \citenamefont {Passerini}}]{tboyle}%
  \BibitemOpen
  \bibfield  {author} {\bibinfo {author} {\bibfnamefont {G.}~\bibnamefont
  {Coccia}}, \bibinfo {author} {\bibfnamefont {G.~D.}\ \bibnamefont {Nicola}},
  \bibinfo {author} {\bibfnamefont {S.}~\bibnamefont {Tomassetti}}, \bibinfo
  {author} {\bibfnamefont {M.}~\bibnamefont {Pierantozzi}},\ and\ \bibinfo
  {author} {\bibfnamefont {G.}~\bibnamefont {Passerini}},\ }\bibfield  {title}
  {\bibinfo {title} {Determination of the {B}oyle temperature of pure gases
  using artificial neural networks},\ }\href
  {https://doi.org/https://doi.org/10.1016/j.fluid.2019.04.003} {\bibfield
  {journal} {\bibinfo  {journal} {Fluid Phase Equilibr.}\ }\textbf {\bibinfo
  {volume} {493}},\ \bibinfo {pages} {36 } (\bibinfo {year}
  {2019})}\BibitemShut {NoStop}%
\bibitem [{\citenamefont {{S. Udry}}\ \emph {et~al.}(2007)\citenamefont {{S.
  Udry}}, \citenamefont {{X. Bonfils}}, \citenamefont {{X. Delfosse}},
  \citenamefont {{T. Forveille}}, \citenamefont {{M. Mayor}}, \citenamefont
  {{C. Perrier}}, \citenamefont {{F. Bouchy}}, \citenamefont {{C. Lovis}},
  \citenamefont {{F. Pepe}}, \citenamefont {{D. Queloz}},\ and\ \citenamefont
  {{J.-L. Bertaux}}}]{Udry}%
  \BibitemOpen
  \bibfield  {author} {\bibinfo {author} {\bibnamefont {{S. Udry}}}, \bibinfo
  {author} {\bibnamefont {{X. Bonfils}}}, \bibinfo {author} {\bibnamefont {{X.
  Delfosse}}}, \bibinfo {author} {\bibnamefont {{T. Forveille}}}, \bibinfo
  {author} {\bibnamefont {{M. Mayor}}}, \bibinfo {author} {\bibnamefont {{C.
  Perrier}}}, \bibinfo {author} {\bibnamefont {{F. Bouchy}}}, \bibinfo {author}
  {\bibnamefont {{C. Lovis}}}, \bibinfo {author} {\bibnamefont {{F. Pepe}}},
  \bibinfo {author} {\bibnamefont {{D. Queloz}}},\ and\ \bibinfo {author}
  {\bibnamefont {{J.-L. Bertaux}}},\ }\bibfield  {title} {\bibinfo {title} {The
  {HARPS} search for southern extra-solar planets. {XI}. {S}uper-{E}arths (5
  and 8 {M}$_\oplus$) in a 3-planet system},\ }\href
  {https://doi.org/10.1051/0004-6361:20077612} {\bibfield  {journal} {\bibinfo
  {journal} {A\&A}\ }\textbf {\bibinfo {volume} {469}},\ \bibinfo {pages} {L43}
  (\bibinfo {year} {2007})}\BibitemShut {NoStop}%
\bibitem [{\citenamefont {Dymond}\ \emph {et~al.}(2003)\citenamefont {Dymond},
  \citenamefont {Marsh},\ and\ \citenamefont {Wilhoit}}]{Dy2003mix}%
  \BibitemOpen
  \bibfield  {author} {\bibinfo {author} {\bibfnamefont {J.~D.}\ \bibnamefont
  {Dymond}}, \bibinfo {author} {\bibfnamefont {K.~N.}\ \bibnamefont {Marsh}},\
  and\ \bibinfo {author} {\bibfnamefont {R.~C.}\ \bibnamefont {Wilhoit}},\
  }\href@noop {} {\emph {\bibinfo {title} {Virial coefficients of pure gases
  and mixtures}}},\ Vol.~\bibinfo {volume} {21}\ (\bibinfo  {publisher}
  {Springer Landord-Bornstein},\ \bibinfo {year} {2003})\BibitemShut {NoStop}%
\end{thebibliography}%

\end{document}